\documentclass[12pt]{article}
\usepackage{enumerate,amssymb}
\renewcommand{\a}{\alpha}
\renewcommand{\b}{\beta}
\newcommand{\g}{\gamma}           
\renewcommand{\d}{\delta}

\newcommand{\ka}{\kappa}

\newcommand{\la}{\lambda}        
\newcommand{\m}{\mu}
\newcommand{\n}{\nu}

\newcommand{\s}{\sigma}           
         
\newcommand{\f}{{\phi}}

\newcommand{\eps}{{\epsilon}}

\newcommand{\be}{\begin{equation}}
\newcommand{\ee}{\end{equation}}
\newcommand{\eqn}[1]{\label{#1}\end{equation}}

\newcommand{\bea}{\begin{eqnarray}}
\newcommand{\eea}{\end{eqnarray}}
\newcommand{\eqan}[1]{\label{#1}\end{eqnarray}}

\newcommand{\ba}{\begin{array}}
\newcommand{\ea}{\end{array}}

\newcommand{\nn}{\nonumber}

\begin{document}

\begin{center}
{\bf  String Corrections To The Riemann Curvature Tensor}\\[14mm]

S. Bellucci\\

{\it INFN-Laboratori Nazionali di Frascati\\
Via E. Fermi 40, 00044 Frascati, Italy\\ mailto: bellucci@lnf.infn.it}\\[6mm]

D. O'Reilly\\

{\it Department of Physical Sciences,\\Embry-Riddle Aeronautical University\\
600 Clyde Morris Blvd.,\\Daytona Beach, FL 32114-3900, USA\\ mailto: doreilly@gc.cuny.edu }\\[6mm]
oo

\end{center}
\vbox{\vspace{3mm}}

\begin{abstract}  The string corrections to the Riemann Curvature tensor
are found to first order in the string slope parameter, here proportional to $\g$. This
is done for D=10 supergravity, the presumed low energy limit of
string theory. We follow the perturbative approach. We also
simplify a crucial result in our previous solution.

 \vbox{\vspace{1mm}}

\end{abstract}

\newpage

\section{Introduction}
In the past we have studied ten dimensional supergravity in
superspace, [1,7]. We followed the perturbative approach of Gates et
al. [2,3,5]. This group found a consistent solution to
the Bianchi identities in superspace, thereby constructing a
manifestly supersymmetric theory, achieving this to first order in $\g$,
[5]. The theory yields a candidate for the low energy limit of
string theory to that order. An alternative approach, given in [6] and [4], is the
the Italian School approach. Furthermore, in [4], deformation techniques were used to study the torsion sector alone. The process to find a second order solution via the perturbative approach took many years. It was
required to find the so called X tensor discussed in [2], and then
finding a consistent set of solutions to the Bianchi identities in
superspace in the torsion, curvature and H sectors at dimension
zero through $\frac{3}{2}$. This was achieved in [1] and [7]  to
second order. Here we modify known results and find the string
corrected Riemann curvature tensor. In particular we simplify a crucial result found in  [7].
Our formalism is well discussed in [1] and [7]. However the following brief outline will
help to make this paper self contained. The Bianchi identities in
superspace are as usual,
\bea[[\nabla_{[A},\nabla_{B}\},\nabla_{C)}\}=0\eea
Where \bea
[\nabla_{A},\nabla_{B}\}=T_{AB}{}^{C}\nabla_{C}+\frac{1}{2}R_{AB}{}^{de}M_{ed}+iF_{AB}^{I}t_{I}
\eea
 The $t_{I}$ are the generators of the Yang-Mills gauge group. In order to illustrate the perturbative approach, we have the following results.
The $G$ tensor obeys
 \bea \nabla_{[A|}G_{|BCD)}-T_{[AB|}{}^E
G_{E|CD)}=0\eea The Lorentz Chern Simons Superform obeys
 \bea \nabla_{[A|}Q_{|BCD)}-T_{[AB|}{}^E
Q_{E|CD)}=R_{[AB|ef}R_{|CD)}{}^{ef}\eea
The H tensor is defined as
\bea H_{ADG}= G_{ADG}-\g Q_{ADG}- \b Y_{ADG} \eea
 Here $Y_{ADG}$ is the Yang Mills Superform, and $\g$ is proportional to the string
slope parameter. $Q_{ADG}$ is the Lorentz Chern-Simons Superform.
Here we set $\b=0 $. Hence $H_{BCD}$ obeys the identity \bea
\nabla_{[A|}H_{|BCD)}-T_{[AB|}{}^E H_{E|CD)}=-\g
R_{[AB|ef}R_{|CD)}{}^{ef}\eea This is the basis of the
perturbative approach as we now can solve for torsions and curvatures as well as H sector tensors at successively higher orders in $\g$.  That is, we first find solutions to (6). We then
proceed to find mutually consistent solutions in the torsion and
curvature sectors. All of the required identities are listed in [1]. We write the
order of the tensors in superscript as follows: \bea
R_{ABde}=R^{(0)}{}_{ABde}+R^{(1)}{}_{ABde}+R^{(2)}{}_{ABde}+...~~~~~~\nn\\
T_{AD}{}^{G}=T^{(0)}{}_{AD}{}^{G}+T^{(1)}{}_{AD}{}^{G}+T^{(2)}{}_{AD}{}^{G}+...~~~~~~~\eea
In our previous work, [1], we found a candidate for the so called X
tensor as discussed in [2]. This along with other techniques, enabled a solution to be found to second order. We proposed the following second order modification to the dimension zero torsion.
\bea T^{(2)}{}_{\a\b}{}^{g}=-\s^{pqref}{}_{\a\b}X_{pqref}{}^{g}=-\frac{i\g}{6}\s^{pqref}{}_{\a\b}
H^{(0)}{}^{g}{}_{ef}A^{(1)}{}_{pqr}\eea We maintained the
definition of the supercurrent as used in [5], \bea A^{(1)}{}_{gef}=i\g \s_{gef \epsilon \tau}T^{mn
\epsilon}T_{mn}{}^{\tau} \eea.
\section{The Basic Analyses}
The object $ R_{abmn}$ in ten dimensional superspace possesses all
of the symmetries and properties of the Riemann curvature tensor.
Hence, we will assume that it reduces to the Riemann tensor in ten dimensions and to the usual four dimensional tensor after dimensional reduction. We have from equation (1), the Bianchi
identity \bea [[\nabla_{[\a},\nabla_{\b}\},\nabla_{a)}\}=0\eea This generates the result
\bea T_{\g[k|}{}^{\la}T_{\la|l]}{}^{\tau}+
T_{\g[k}{}^{g}T_{g|l]}{}^{\tau}+T_{kl}{}^{\la}T_{\la
\g}{}^{\tau}+T_{kl}{}^{g}T_{g\g}{}^{\tau}-\nabla_{[k|}T_{|l]\g}{}^{\tau}
-R_{kl\g}{}^{\tau}-\nabla_{\g}T_{kl}{}^{\tau}=0\eea We will use
this, along with some other results, to calculate the corrections
to $ R_{abmn}$. The required torsions to first order, which we select from the complete set listed in the appendix, are as follows:
\bea T^{(0)}{}_{\a\b}{}^{g}=i\s_{\a\b}{}^{g};~~T^{(1)}{}_{\a\b}{}^{g}=0;~~T^{(0)}{}_{\a\b}{}^{\la}=-[\d_{(\a|}{}^{\la}\d_{|\b)}{}^{\d}+\s^{g}{}_{\a\b}\s_{g}{}^{\la
\d}]\chi_{\d}\nn\\
T^{(1)}{}_{\a\b}{}^{\la}=0;~~ T_{abc}=-2L_{abc};~~
T^{(0)}{}_{\g gd}=T^{(1)}{}_{\g gd}=0~~~~~\eea These results were first found in [5]. We also maintain the
following conventional constraint, at first order, as used in [5].
(This was later modified at second order,[1].) \bea T^{(1)}{}_{\a a}{}^{\b}=
-\frac{1}{48}\s_{a}{}_{\a\la}\s^{pqr}{}^{\b\la}A_{pqr} \eea
 We recall that
\bea
R_{kl\g}{}^{\tau}=\frac{1}{4}R_{klmn}\s^{mn}{}_{\g}{}^{\tau}\eea
And we also have the identity \bea
\s^{mn}{}_{\g}{}^{\tau}\s_{rs}{}_{\tau}{}^{\g}=-16\d^{m}{}_{[r|}\d^{n}{}_{|s]}
\eea Hence we arrive at our basic equation for $R_{klmn}$.\bea
R_{klmn}=-\frac{1}{8}\s_{mn}{}_{\tau}{}^{\g}\{T_{\g[k|}{}^{\la}T_{\la|l]}{}^{\tau}+
T_{\g[k}{}^{g}T_{g|l]}{}^{\tau}+T_{kl}{}^{\la}T_{\la
\g}{}^{\tau}+T_{kl}{}^{g}T_{g\g}{}^{\tau}-\nabla_{[k|}T_{|l]\g}{}^{\tau}
-\nabla_{\g}T_{kl}{}^{\tau}\}\nn\\\eea This is valid to all
orders. Substituting the torsions in (12) into (16) gives \bea
R^{(0)}{}_{klmn}=-\frac{1}{8}\s_{mn}{}_{\tau}{}^{\g}\{T_{kl}{}^{\la}T_{\la
\g}{}^{\tau} -\nabla_{\g}T_{kl}{}^{\tau}|^{Order(0)}\}\eea \bea
R^{(1)}{}_{klmn}=-\frac{1}{8}\s_{mn}{}_{\tau}{}^{\g}\{T_{kl}{}^{g}T_{g\g}{}^{\tau}-\nabla_{[k|}T_{|l]\g}{}^{\tau}
-\nabla_{\g}T_{kl}{}^{\tau}|^{Order(1)}\}\nn\\\eea In reference
[7], in order to close the curvature identity at dimension
$\frac{3}{2}$, we required \bea
R^{(1)}{}_{kl\g}{}^{\tau}=+\frac{\g}{100}T{}_{kl}{}^{\tau}[T_{mn}{}^{\la}R^{(0)}{}_{\g
\la}{}^{mn}
+T^{(0)}{}_{mn}{}^{g}R^{(0)}{}_{\g}{}_{g}{}^{mn}-\nabla_{\g}R^{(0)}{}_{mn}{}^{mn}]\nn\\
+\frac{1}{48}[T^{(0)}{}_{klg}\s^{g}{}_{\g\la}\s^{rst \la \tau}A^{(1)}{}_{rst}
+2\s_{[k|\g \la}\s^{rst}{}^{\la\tau}\nabla_{|l]}A^{(1)}{}_{rst}] \eea
However we also have the Bianchi identity
\bea T_{\a [a|}{}^{X}R_{X}{}_{|b]d e}
+T_{ab}{}^{X}R_{X}{}_{\a de}-\nabla_{[a|}R_{|b]\a d e}-\nabla_{\a}R_{abde}=0 \eea
Using (20) in (19) gives
\bea
R^{(1)}{}_{kl\g}{}^{\tau}=+\frac{i\g}{50}T_{kl}{}^{\tau}[\s^{[m|}{}_{\g\la}\nabla_{m}T^{|n]}_{n}{}^{\la}]
+\frac{1}{48}[T^{(0)}{}_{klg}\s^{g}{}_{\g\la}\s^{rst \la \tau}A^{(1)}{}_{rst}
+2\s_{[k|\g \la}\s^{rst}{}^{\la\tau}\nabla_{|l]}A^{(1)}{}_{rst}] \nn\\\eea
The first term on the RHS of (19) and (21) is therefore shown to
vanish. This now simplifies the result first found in [7], and hence greatly simplifies the calculation at third order. We find that (18) and (20) become \bea R^{(1)}{}_{kl\g}{}^{\tau}=
+\frac{1}{48}[T^{(0)}{}_{klg}\s^{g}{}_{\g\la}\s^{rst \la
\tau}A^{(1)}{}_{rst} +2\s_{[k|\g
\la}\s^{rst}{}^{\la\tau}\nabla_{|l]}A^{(1)}{}_{rst}] \eea The use
of the identity \bea
\s^{pqr~\eps\tau}\s_{pqr~\m\n}=-48\d^{\eps}{}_{[\m|}\d^{\tau}{}_{|\n]}
\eea as well as equation (9), simplifies (22) to \bea
R^{(1)}{}_{klab}=\frac{i\g}{4}\s_{ab}{}_{\tau}{}^{\g}\s^{g}{}_{\g\la}T_{klg}T_{mn}{}^{\la}T^{mn~\tau}
+\frac{i\g}{2}\s_{ab}{}_{\tau}{}^{\g}\s_{[k|\g\la}T_{mn}{}^{\la}\nabla_{|l]}T^{mn\tau}\eea
Hence, we have the first order corrections to the Riemann curvature
tensor as required, or main result.
\section{Conclusion}

As is well known, the theory of strings, in a striking difference
with respect to particles, involves a length scale, i.e. the
string slope parameter, here proportional to $\g$. Supersymmetry
and string theory have been realized to play a cooperative
proactive role in several instances/scenarios, e.g. the landscape
of supersymmetric compactifications, the AdS/CFT correspondence
and the understanding of N = 4 supersymmetric Yang-Mills
(including exact results). Supergravity can be viewed as the
low-energy limit of string theory. The latter possesses two
expansion parameters, i.e. the string length, proportional to
$\g$, which determines the corrections to supergravity results,
and the string coupling constant, entering the string loop
expansion. String theory naturally lives in ten dimensions, which
naturally led to the revival of the Kaluza-Klein notion of
compactification, with the well-known problematic implications in
terms of moduli stabilization. The cosmological reflections of the
theory led to the model by Kachru, Kallosh, Linde, Trivedi for the
construction of supersymmetric compactifications [8]. In spite of
the tremendous success of the model, a point deserves careful
attention, connected to the loss of control of the ten dimensional
solution and the need to involve classical and quantum effects in
moduli stabilization. The latter may lead one to address the issue
of the reliability of the treatment of non-perturbative
corrections, connected to the spacetime being $AdS_4$. The
fundamental reason for the difficulty in finding an explicit and
simple de Sitter solution is inherent in the loss of
supersymmetry, which requires solving the full supergravity
equations of motion, rather than the much less involved task of
imposing supersymmetry and solving Bianchi identities.

Based upon considerations such as the abovementioned ones, we
revived the issue of providing a manifestly supersymmetric
formulation of $\g$ parturbed D=10, N=1 supergravity with fully
consistent component-level symmetry transformation laws for the
theory, to the relevant order in the $\g$ perturbation. Based upon
the results obtained in the present work, we can in a future work
construct the corrected Ricci tensor and the corrected Scalar
Curvature.

We simplified our previously published solution to second order by
reducing (19) to (22). We then found the superspace expression for
the string corrected Riemann curvature tensor, as it arises in ten
dimensional perturbed supergravity, following the procedure
outlined in equations (3) through (6). We now recall that the
effective action can be written as an expansion in the parameter
$\g$.

 \bea S_{eff}=\frac{1}{\ka^{2}}\int
d^{10}x
e^{(-1)}[\emph{L}_{(0)}~+~\sum_{n=1}^{n=\infty}(\gamma')^{n}\emph{L}_{(n)}]
\eea Let us suppose in the minimal scenario we consider the non
supersymmetric low energy case of a string corrected General
Relativity, which does not involve higher derivative gravity
terms, nor the Gauss Bonnet term expected to arise from string
corrected gravity, hence limiting our consideration simply to the
Einstein Hilbert action. We might like to ask what minimal
contributions result from constructing a curvature scalar using
(24). We now note that we can indeed proceed to higher orders by
using the  torsions listed in appendix, in conjunction with
equation (16). \setcounter{equation}0
\section*{Acknowledgements}
This work was partially supported by the ERC Advanced
Grant no. 226455, \textit{``Supersymmetry, Quantum Gravity and Gauge Fields''%
} (\textit{SUPERFIELDS}).

\section{Appendix: Main Results to Second
Order} \bea H_{\a\b\g}=0 + order(\g^2) \eea \bea
H^{(0)~g}{}_{\a\b}= \frac{i}{2}\s^{g}{}_{\a\b};~~~~
H^{(1)~g}{}_{\a\b }=
4i\g\s^{p}{}_{\a\b}H^{(0)}{}_{pmn}H^{(0)}{}^{gmn} \eea \bea
H^{(2)}{}_{\a\b g}=
+\s_{\a\b}{}^{g}[{8i\g}H^{(0)}{}_{def}[L^{(1)}{}_{g}{}^{ef}-\frac{1}{8}A^{(1)}{}_{g}{}^{ef}]-
\frac{i\g}{12}\s^{pqref}{}_{\a\b}H^{(0)}{}_{def}A^{(1)}{}_{pqr}
\eea Where as reported in [5]

\bea L_{gef}= H^{(0)}{}_{gef}+\g\{
R^{(0)}{}^{mn}{}_{[ef|}H^{(0)}{}_{|g]mn}
+ R^{(0)}{}_{[ef|}{}^{mn}H^{(0)}{}_{|g]mn}-\nn\\
\frac{8}{3}H^{(0)}{}_{pm
[e|}H^{(0)}{}^{p}{}_{|f|n}H^{(0)}{}_{|g]}{}^{mn}\} \eea
\bea
T_{abc}=-2L_{abc}
\eea
Its spinor derivative is required to be, [5],
\bea \nabla_{\a}L_{bcd}{}^{(0)}=\frac{i}{4}\s_{[b|\a\b}T_{|cd]}{}^{\b};~~~~\nabla_{\a}L_{bcd}{}^{(1)}=i\g\s_{[b|\a\b}
T_{kl}{}^{\b}R^{kl}{}_{|cd]}=R^{(1)}{}_{\a bde}\eea
\bea T^{(0)}{}_{\a\b}{}^{g}= i\s_{\a\b}{}^{g}=2H^{(0)}{}^{g}{}_{\a\b};~~~
T^{(1)}{}_{\a\b}{}^{g}=0;~~~
T^{(2)}{}_{\a\b}{}^{g}=-\frac{i\g}{6}\s^{pqref}{}_{\a\b}
H^{(0)}{}^{g}{}_{ef}A^{(1)}{}_{pqr}\eea
\bea
T^{(0)}{}_{\a b}{}^{\g}=0;~~~
T^{(1)}{}_{\a b}{}^{\g} =
 -\frac{1}{48}\s_{b \a \f}\s^{pqr}{}^{\f\g}A_{pqr};~~~
T^{(2)}{}_{\g g}{}^{\la}= 2\g\Omega^{(1)}{}_{\g gef}T^{ef}{}^{\la}
\eea
Where
\bea
\Omega^{(1)}{}_{gef}= L^{(1)}{}_{gef}-\frac{1}{4}A^{(1)}{}_{gef};~~ \Omega^{(1)}{}_{\a gef} = \nabla_{\g}\{L^{(1)}{}_{gef}-\frac{1}{4}A^{(1)}{}_{gef}\} \eea
\bea T^{(0)}{}_{\a\b}{}^{\g}=-[\d_{(\a|}{}^{\g}\d_{|\b)}{}^{\d}+\s^{g}{}_{\a\b}\s_{g}{}^{\g
\d}]\chi_{\d};~~ T^{(1)}{}_{\a\b}{}^{\g}=0\eea
\bea
T^{(2)}{}_{\a\b}{}^{\la}=-\frac{i\g}{12}\s^{pqref}{}_{\a\b}A^{(1)}{}_{pqr}T_{ef}{}^{\la}
\eea
\bea T^{(0)}{}_{\g gd}=T^{(1)}{}_{ \g gd}=0\eea

\bea\s^{g}{}_{(\a\b|}T^{(2)}{}_{|\g)gd}
=4\g\s^{g}{}_{(\a\b|}\Omega_{|\g) gef}H^{(0)}{}_{d}{}^{ef}-
\frac{i\g}{6}\s^{g}{}_{(\a\b|}\s^{pqre}{}_{g|\g)\f}A^{(1)}{}_{pqr}T^{(0)}{}_{de}{}^{\f}\nn\\
=\s^{g}{}_{(\a\b|}X_{|\g) gd}~~~~~~~~~~~~~~~~~~~~~~~~~~~~~~~~~~\eea
Defining
\bea \hat{O}^{ab  \b}_{gd \g}=[\frac{1}{2}\d_{[g}{}^{a}\d_{d]}{}^{b}\d_{\g}{}^{\b}~
-~\frac{1}{12}\eta^{ab}\s_{gd\g}{}^{\b}~+\frac{1}{24}\d_{[g}{}^{(a}\s_{d]}{}^{b)}_{\g}{}^{\b}]
\eea
Therefore
\bea
T^{(2)}{}_{\g gd}= \hat{O}^{ab  \b}_{gd \g}[X_{\b ab}]
\eea
\bea R^{(0)}{}_{\a\b de}=-2i\s^{g}{}_{\a\b}H^{(0)}{}_{gde};~R^{(1)}{}_{\a\b de}=-2i\s^{g}{}_{\a\b}[L^{(1)}{}_{gde}
-\frac{1}{8}A^{(1)}{}_{g}{}^{ef}]+\frac{i}{24}\s^{pqrde}{}_{\a\b}A_{pqr}\eea
\bea R^{(2)}{}_{ \a\b de}
 =-\frac{i\g}{12}\s^{pqref}{}_{\a\b}A^{(1)}{}_{pqr}R_{ef de};~~
 R^{(0)}{}_{\a bde}= -i\s_{[d|}{}_{\a \f}T_{b |e]}{}^{f}\eea
\bea R^{(1)}{}_{\a bde}= i\g\s_{[d|}{}_{\a \f}T_{kl}{}^{\f}R^{kl}{}_{|de]};~~~
R^{(2)}{}_{\g gde}=2\g \Omega^{(1)}{}_{\g gmn}R^{(0)mn}{}_{de}
\eea

\end{document}